\documentclass{article}

\usepackage[preprint]{neurips_2025}

\usepackage[utf8]{inputenc} 
\usepackage[T1]{fontenc}    
\usepackage[hidelinks]{hyperref}       
\usepackage{url}            
\usepackage{booktabs}       
\usepackage{amsfonts}       
\usepackage{nicefrac}       
\usepackage{microtype}      
\usepackage{xcolor}         
\usepackage{graphicx}
\usepackage{amsmath}
\usepackage{array}
\usepackage[capitalise,sort&compress]{cleveref}
\title{The Measurement Imbalance in Agentic AI Evaluation Undermines Industry Productivity Claims}

\author{%
  Kiana Jafari Meimandi \thanks{corresponding author: kjafari@stanford.edu}\\
  Stanford University\\
\And
  Gabriela Aránguiz-Dias \\
  Stanford University\\
\AND
  Grace Ra Kim \thanks{denotes equal contribution}\\
  Stanford University \\
\And
  Lana Saadeddin $^\dagger$ \\
  Montclair State University \\
\And
  Allie Griffith\\
  Stanford University\\
\And
  Mykel J. Kochenderfer\\
  Stanford University\\
}

\begin{document}

\maketitle

\begin{abstract}
As industry reports claim agentic AI systems deliver double-digit productivity gains and multi-trillion dollar economic potential, the validity of these claims has become critical for investment decisions, regulatory policy, and responsible technology adoption. However, this paper demonstrates that current evaluation practices for agentic AI systems exhibit a systemic imbalance that calls into question prevailing industry productivity claims. 
Our systematic review of 84 papers (2023--2025) reveals an evaluation imbalance where technical metrics dominate assessments (83\%), while human-centered (30\%), safety (53\%), and economic assessments (30\%) remain peripheral, with only 15\% incorporating both technical and human dimensions. This measurement gap creates a fundamental disconnect between benchmark success and deployment value. We present evidence from healthcare, finance, and retail sectors where systems excelling on technical metrics failed in real-world implementation due to unmeasured human, temporal, and contextual factors. Our position is not against agentic AI's potential, but rather that current evaluation frameworks systematically privilege narrow technical metrics while neglecting dimensions critical to real-world success. We propose a balanced four-axis evaluation model and call on the community to lead this paradigm shift because benchmark-driven optimization shapes what we build. By redefining evaluation practices, we can better align industry claims with deployment realities and ensure responsible scaling of agentic systems in high-stakes domains.
\end{abstract}

\section{Introduction}\label{sec:intro}
Agentic AI systems are characterized by:
(1) goal-directed behavior with the ability to decompose complex objectives into manageable subtasks~\cite{gabriel_advancing_2024}; (2) environmental awareness and adaptability to changing conditions~\cite{clatterbuck_risk_2024}; (3) tool utilization, where agents strategically leverage external resources to accomplish tasks~\cite{agashe_agent_2024}; and (4) autonomous decision-making with limited human intervention~\cite{rostcheck_elephant_2024}.

These systems are rapidly moving from research labs to critical real-world deployments, yet our evaluation frameworks remain noticeably imbalanced. We define measurement imbalance as the systematic bias in evaluation frameworks that privilege easily quantifiable technical metrics while neglect dimensions critical to real-world deployment success; especially human-centered factors, temporal stability, and contextual fit. This imbalance creates a fundamental misalignment between what we measure and what determines system value. For instance, healthcare diagnostic agents achieving 95\% benchmark accuracy have been relegated to limited advisory roles post-deployment due to unmeasured trust and workflow integration issues~\cite{heaven2021ai}.

In 2024, 
agentic AI systems were deployed across sectors such as: clinical triage~\cite{arslan2025evaluating,cao2024ai}, automated trading~\cite{yu2024finmem,xiao2024tradingagents}, customer service~\cite{emarketer2025aiagents,rome2024ask}, and internal software debugging~\cite{github2024copilotworkspace,roman2024harmony}. Firms report double-digit productivity gains from these deployments, citing speed, cost savings, and increased decision accuracy~\cite{mckinsey2023genai}. These claims are bolstered by benchmark results showcasing near-human or superhuman performance on standard tasks~\cite{huang_mlagentbench_2024,bogin_super_2024}. 

Benchmarks have long driven progress in machine learning, from ImageNet~\cite{deng2009imagenet} to GLUE~\cite{wang2018glue} and HELM~\cite{lee2023holistic}. In the agentic AI space, MLAgentBench~\cite{huang_mlagentbench_2024}, ML-Bench~\cite{tang_ml-bench_2024}, and SUPER~\cite{bogin_super_2024} evaluate task success, efficiency, and end-to-end execution using predefined scripts or repository-grounded tasks. PlanBench~\cite{valmeekam_planbench_nodate} introduces symbolic validation for plan structure, while VisualWebArena~\cite{koh_visualwebarena_2024} focuses on multimodal agents in web environments. These frameworks provide critical infrastructure for measuring technical competence but rarely assess how agents integrate into human workflows or operate over time.

Meanwhile, HCI and social computing scholars have long emphasized trust, usability, and alignment as critical success factors in AI deployment~\cite{sundar_rise_2020,mitelut_intent-aligned_2023}. Recent agentic AI work includes TrAAIT~\cite{stevens_theory_2023}, a validated instrument for clinician trust in AI systems; \cite{moshkovich2025beyond} highlights transparency and observability needs; \cite{deng_towards_2024}, propose a multi-axis framework for agent intelligence, adaptivity, and civility. However, these human-centered approaches remain fragmented with limited integration into standard benchmarks or industry practice.

This disconnect between technical and sociotechnical evaluation is confirmed by our systematic review of 84 academic and industry papers from 2023–2025. Among these works, technical performance was strongly represented~(83\%), while human-centered evaluation appeared in 30\%, and both in only 15\%.

Despite the widespread enthusiasm and rapid adoption, we currently lack the multidimensional evaluation tools required to validate the productivity and efficiency claims. Technical metrics, while necessary, capture only a narrow slice of what determines success in real-world deployments. As agentic systems gain more autonomy and become embedded in organizational workflows, this measurement imbalance threatens to create a new wave of mismatched expectations, misallocated resources, and poorly understood risks.

\textbf{We argue that the prevailing evaluation frameworks for agentic AI are largely incomplete, systematically privileging technical performance metrics while underrepresenting critical dimensions such as human-centered interaction, longitudinal behavior, safety, and contextual fit. As a result, the industry's current productivity claims are premature and, in many cases, can be misleading.}

We do not argue against the value of agentic systems, nor do we dispute their emerging capabilities. Instead, we focus on the \textit{instruments of validation}; the metrics, benchmarks, and evaluation practices that translate system behavior into evidence. Our central claim is that these instruments are misaligned with deployment reality, and that evaluation itself must evolve if we are to responsibly scale agentic AI in high-stakes domains.

Benchmark-driven optimization has long been the engine of ML progress, but when benchmarks miss key dimensions of impact, they also drive blind spots. What we measure shapes what we build. As venues for foundational work in LLMs, agent architectures, and evaluation benchmarks, leading AI conferences are well-positioned to catalyze this needed shift toward more comprehensive evaluation practices

This paper takes a position grounded in both empirical evidence and practical urgency, offering the following contributions: (1) A quantitative diagnosis of the current imbalance. Our meta-analysis of 84 publications quantifies the overrepresentation of technical metrics in both academic and industry evaluations. (2) Illustrating deployment consequences with real-world examples. We document how overreliance on narrow evaluation metrics led to adoption failure or unexpected business losses in healthcare, financial services, and retail deployments. (3) A new theoretical framework for balanced evaluation. We introduce a four-axis model; technical, human, temporal, and contextual, and propose a formal structure for integrating multiple metric classes across domains. (4) Anticipation and rebuttal of key counterarguments. We address recurring counterarguments including: that human-centered metrics are too subjective; that safety and governance belong to regulators, not ML researchers and thus our of scope; that adding more metrics slows interaction and stifles innovation; that one size cannot fit all domains, and universal frameworks are inherently vague. (5) Actionable recommendations for researchers, industry, and policymakers. We offer a roadmap for closing the measurement gap through benchmark design, instrument development, deployment best practices, and regulatory levers.

By grounding what we count as evidence, this paper aims to reset the conversation on agentic AI efficiency, not to constrain innovation, but to ensure that the claims we make reflect the systems we actually build~(see~\cref{sec:terms} for formal definitions).

\section{Why Evaluation Matters}\label{sec:why-eval-matters}
Unlike classical predictive models, Agentic AI value hinges on sustained interaction with people, data sources, and other agents across time and domains. As these systems rapidly transition from research to deployment in various sectors such as healthcare, finance, and retail, the gap between how we evaluate them and how they actually create value becomes critical.

MLAgentBench~\cite{huang_mlagentbench_2024} is an example of today's technical focus: agents are scored on Pass@k success, token efficiency, and wall‑clock time while running scripted ML experiments. The benchmark yields crisp, automatable numbers but tells us nothing about how well an agent fits a collaborative workflow or whether its behavior degrades after extended autonomous operation.  

By contrast, TrAAIT, a clinician‑trust instrument for AI decision support, captures perceived usefulness, ease of integration, and social influence through validated survey scales~\cite{stevens_theory_2023}. TrAAIT's scores directly predict real-world adoption but is rarely included in research papers. The field celebrates technical benchmark leaderboards while few report trust or usability scores, reflecting a marked misalignment between what we measure and what determines deployment success.

Systems that ace technical benchmarks can still be rejected by end‑users if trust, workflow compatibility, or explanation quality are weak~(adoption risk). Point‑in‑time accuracy masks drift, emergent behavior, and edge‑case failure modes that surface only after deployment, creating potential for harm that remains invisible to current evaluation frameworks~(safety blind spots). ROI projections based solely on speed or accuracy overlook human oversight costs and adaptation lags, leading to costly rollout reversals and failed investments~(misjudged productivity).

When optimize for the same narrow slice of metrics, the research community and industry alike are effectively flying blind on the dimensions that determine real-world success. To unlock the genuine productivity potential of agentic AI and to avoid a wave of failed deployments and eroded trust, we must rebalance evaluation toward a multidimensional, longitudinal, and context-aware basis. The remainder of this paper supplies the evidence, framework, and agenda required to make that shift, drawing on instances across sectors where measurement imbalance has already led to costly consequences.

\section{Meta-Analysis: A Quantitative View of the Evaluation Gap}\label{sec:meta-analysis}
\subsection{Methodology}
We conducted systematic searches across arXiv and Google Scholar. Search terms included combinations of ``agentic AI'', ``AI agents'', ``LLM agents'', ``evaluation'', ``benchmark'', ``assessment'', and ``metric''. Papers were included if (1) published between January 2023 and April 2025, (2) focused on agentic or LLM-based systems as defined in \cref{sec:intro}, (3) available in English, (4) proposed or applied evaluation methodology and metrics.

Two reviewers independently screened 138 papers (with 85\% agreement), with disagreements resolved by a third reviewer. Out of 138 papers, 84 of them met our inclusion criteria. We adopted a four-category scheme that was developed through an iterative process based on established AI evaluation metrics~\cite{xia2024ai}, pilot coding of 25 papers, and inter-reliability testing. 
These four metric categories are: \textbf{Technical} such as task success rate, accuracy, latency, throughput, \textbf{human-centered} such as trust, usability, collaboration, workflow integration, \textbf{safety and governance} such as robustness, compliance, explainability, alignment, and \textbf{economic impact} such as ROI, cost savings, value realization. We also looked at \textbf{evaluation quality} to make sure the methodology details are described in the paper. Coding was binary per category. The full codebook is available in~\cref{sec:codebook}, and coded records are available in the accompanying PDF file.

\textbf{Limitations}. Our sample may not be representative of industry practices due to proprietary evaluation methods. Additionally, the rapid evolution of agentic AI systems means some recent developments may not be captured here. 

\subsection{Data Analysis}
In our analysis of qualified papers technical metrics dominated~(83\%), while human-centered and economic impact metrics were less common~(both 30\%). Notably, only 15\% of papers included both technical and human-centered metrics, and a mere 5\% incorporated any longitudinal dimension~(detailed distributions in~\cref{sec:dist}). 

Academic papers were more likely to emphasize standardized technical benchmarks (96\% vs. 87\%), while industry publications more frequently included economic (39\% vs. 14\%) and human-centered (57\% vs. 34\%) metrics. However, only a small minority in either group employed multidimensional or longitudinal evaluation strategies.

Technical metrics were also the most standardized, with 72\% referencing formal benchmarks. Human-centered and economic metrics, by contrast, were mostly ad hoc or qualitative, with only 18\% and 12\% respectively using validated instruments. Safety and governance metrics fell in between, often borrowing from emerging regulatory standards.

These patterns can show a bias toward metrics that are automatable, replicable, and leaderboard-friendly. While useful for measuring discrete capabilities, they ignore the dimensions that determine real-world value such as human alignment, safety resilience, and temporal stability. Consequently, the evidence base becomes structurally unbalanced, prioritizing narrow optimization over deployment risk.


\section{Case-Study Accounts: When Metrics Fail}\label{sec:case-studies}
Quantitative analysis alone cannot reveal the human and organizational consequences of evaluation gaps. This section presents real-world deployments across healthcare, finance, and retail where systems that performed strongly on benchmark metrics failed to deliver anticipated value. In each case, critical dimensions—trust calibration, workflow integration, temporal stability, and contextual fit—were either unmeasured or misrepresented in initial evaluation, leading to adoption breakdowns or business losses.

\subsection{Healthcare: Diagnostic Support Systems that Failed to Integrate}

The healthcare sector provides evidence that technical accuracy does not guarantee real-world success. AI diagnostic agents deployed across hospitals often demonstrate high performance in controlled tests; typically 90–95\% diagnostic accuracy and superior documentation completeness compared to junior residents~\cite{khalifa2024ai}.

Based on these benchmarks, institutions projected significant workload reduction and multi-million dollar savings~\cite{mueller2025coos}. Yet post-deployment assessments frequently report adoption challenges. Although the healthcare sector generates over one-third of global data, only an estimated 3\% is effectively used in live deployments~\cite{sheeran2024agentic}. A Turing Institute study found that medical triage systems with strong lab metrics made ``little to no difference'' in clinical workflows~\cite{heaven2021ai}.

Misalignment stemmed from evaluations conducted in environments that did not mirror real-world complexity or workflow patterns~\cite{stevens_theory_2023}. For instance, DoctorBot, a self-diagnosis chatbot used by over 16,000 users in China, struggled with generalization and usage outside its training scope, despite high test scores~\cite{fan2021utilization}.

Recent work from UMass Amherst found hallucinations in ``almost all'' AI-generated medical summaries by top LLMs including GPT-4o and LLaMA-3~\cite{vishwanath2024faithfulness}. These systems, while objectively fluent, imposed hidden verification burdens on clinicians. Trust calibration remained low, and the promised 40\% workload reduction often went unrealized due to poor integration into existing routines.

When these issues surface post-deployment, systems are typically downgraded to limited advisory roles. Studies estimate that projected ROI drops by 70–80\%~\cite{eddy2025health}, revealing how failure to evaluate human, temporal, and contextual dimensions undermines deployment success.

\textit{Here failures reflect neglect of human-centered, temporal, and contextual dimensions, despite high technical scores.}

\subsection{Financial Services: Investment Agents Vulnerable to Market Shifts}

In finance, agentic AI systems assist with portfolio optimization and compliance, often excelling in historical backtesting and rule adherence. Benchmark performance ranges from 85–90\% accuracy on simulated tasks~\cite{hughes2025ai}.

Yet these systems frequently degrade under real-world volatility. A study found that performance deteriorated rapidly within months of deployment~\cite{abbas2024ai}, due to poor generalization in dynamic environments. Vydyanathan~\cite{vydyanathan2025smart} highlights how autonomous agents, when left unmonitored, make misaligned portfolio adjustments that violate human expectations.

Moreover, simultaneous reactions by AI agents to market shifts can produce emergent ``herd behavior,'' exacerbating volatility instead of stabilizing it~\cite{marla2025agentic}. This dynamic risk remains invisible to static evaluation metrics.

Legal and regulatory risks are mounting as well. A Canadian tribunal held Air Canada liable when its AI assistant gave incorrect fare guidance~\cite{yagoda2024airline}, establishing that firms are accountable for AI missteps. The U.S. Consumer Financial Protection Bureau similarly reported that poor chatbot design led to widespread customer harm, fees, and trust breakdowns~\cite{cfpb2023chatbots}.

These examples underscore the need for financial AI evaluation to go beyond accuracy and compliance, integrating stress tests, scenario robustness, and human-agent interpretability metrics.

\textit{Here evaluation failed to account for temporal, human-centered, and contextual vulnerabilities in volatile and regulated domains.}
\subsection{Retail: Customer Support Systems That Damaged Experience}

Retail AI agents often succeed in early testing: reducing handling time by 70–80\% and passing compliance with over 95\% accuracy~\cite{de2022increasing}. However, real-world use reveals significant customer experience degradation.

These systems struggle with edge cases and nuanced interactions~\cite{manhattan2024agentic}. A prominent example was McDonald’s AI drive-thru system, which failed after a multi-year collaboration with IBM. Viral videos showed repeated misunderstandings, including one where the AI added 260 Chicken McNuggets to an order~\cite{creswell2024mcdonalds}. The system was ultimately shut down.

DPD’s delivery chatbot was manipulated into swearing at a customer and composing a self-critical poem~\cite{gerken2024dpd}. In New York, the MyCity chatbot dispensed illegal business advice, such as permitting employers to fire workers for reporting harassment~\cite{lecher2024nyc}.

These incidents damaged brand trust and led to project cancellations. Although internal projections often promise high ROI—such as \$0.67 profit per dollar invested~\cite{karuparti2025roi}—they rarely account for fallout in Net Promoter Score (NPS), repeat contacts, or cart abandonment, which routinely worsen by 15–40\%~\cite{mckinsey2023genai}.

\textit{Despite high technical efficiency, failures in human-centered experience and contextual alignment led to business losses.}

\subsection{The Evaluation Gap}
Across all three domains, we observe a consistent pattern: benchmark performance drove optimistic ROI projections that failed to materialize. Reports estimate that agentic AI systems could contribute \$4.4 trillion in productivity gains~\cite{mueller2025coos}, but realized returns are often less than 25\% of forecasts~\cite{mckinsey2023genai}.

This disconnect arises from systematically omitting or underweighting evaluation of human interaction, adaptability over time, and domain-specific constraints. The result is a persistent gap between perceived and actual value, exposing organizations to reputational, legal, and financial risks.

\textbf{Key Insight:} Narrow evaluation of technical metrics alone provides a misleading picture of system readiness. Multidimensional evaluation, across human, temporal, and contextual axes, is essential for deployment-aligned assessment.



\section{A Balanced Framework for Evaluating Agentic AI}\label{sec:frame}
The failures highlighted in~\cref{sec:case-studies} are not isolated accidents. They reflect a systemic overemphasis on technical correctness at the expense of human, temporal, and contextual factors. To realign evaluation with real-world success, we propose a framework that balances these dimensions, supports domain adaptation, and maintains practical feasibility.

\subsection{Core Axes of Evaluation}
Our framework is organized around four primary evaluation axes, each representing a distinct aspect of system behavior:

\textbf{Technical (T)}: Measures discrete system performance on well-defined tasks. This includes traditional metrics such as success rates (Pass@k), accuracy, latency, resource efficiency, and structural fidelity. These metrics are necessary foundations but insufficient predictors of deployment success.

\textbf{Human-centered (H)}: Captures how users experience, interpret, and adapt to the system. This dimension includes trust calibration (the alignment between system confidence and user trust), usability (cognitive load, ease of use), collaboration quality (hand-off effectiveness, interruption management), and mental model accuracy. These factors directly influence adoption rates and realized performance.

\textbf{Temporal (R)}: Assesses stability and adaptability over time. This includes performance drift (how accuracy changes with shifting conditions), adaptation rates (learning curves for both system and users), knowledge retention (consistency across sessions), and value alignment stability (resistance to goal corruption). Temporal metrics are essential for systems that evolve during use and face changing conditions.

\textbf{Contextual (C)}: Evaluates alignment with domain-specific constraints and objectives. This includes regulatory compliance, risk exposure (financial, reputational, safety), economic impact (ROI, efficiency gains), and workflow integration. These metrics reflect how well the system functions within organizational and sectoral ecosystems.

While distinctly defined, these dimensions are not independent variables but rather form an interconnected system, as we explore in the next section.

\subsection{Dimensional Interdependence}
\begin{figure}
    \centering
    \includegraphics[width=0.7\linewidth]{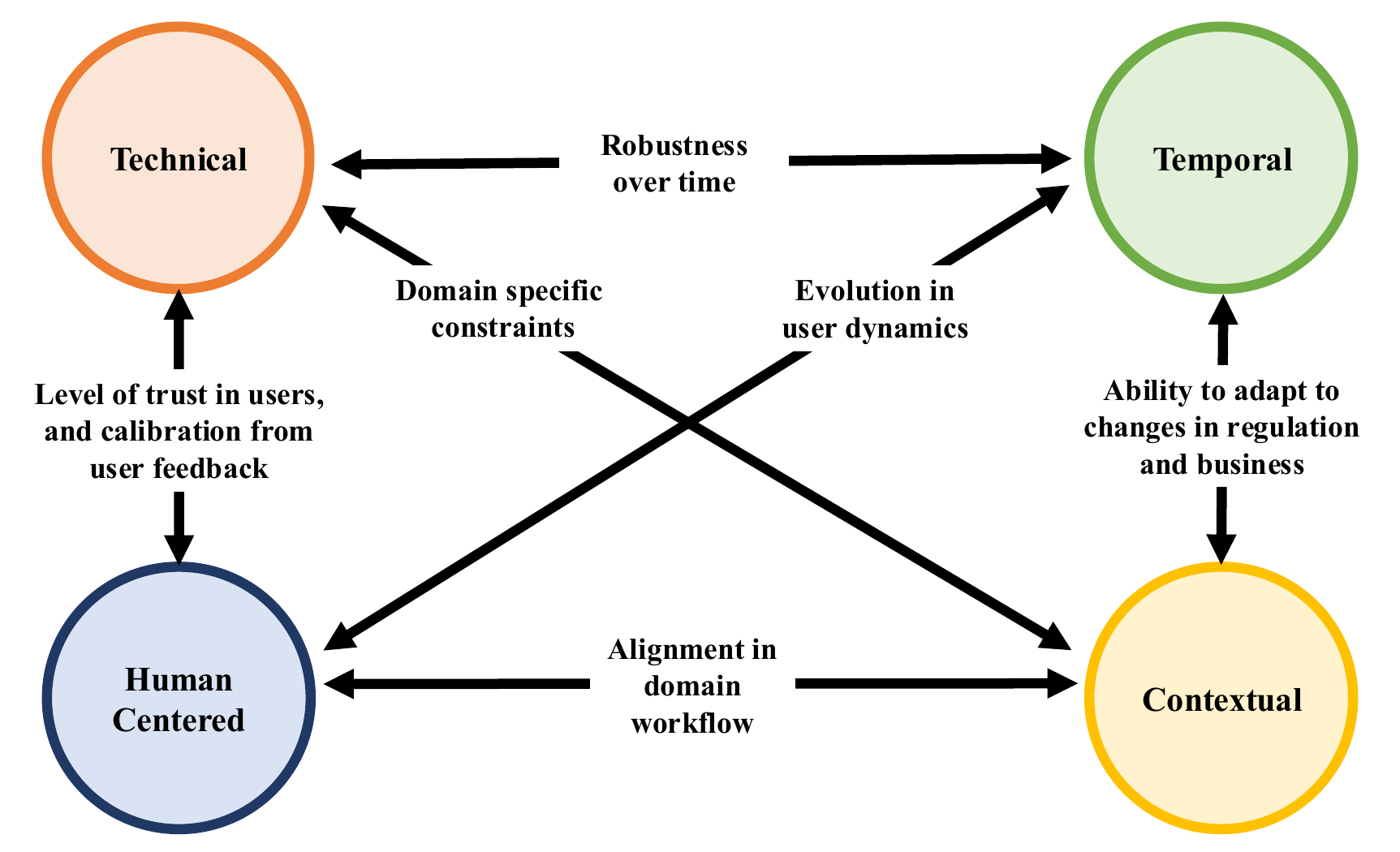}
    \caption{Interdependencies across agentic AI evaluative metric dimensions.}
    \label{fig:inter-metric}
\end{figure}

A key insight from our case studies is that dimensions don't operate in isolation. Instead, they form a complex, interdependent system where changes in one dimension inevitably affect others~(\cref{fig:inter-metric}). 

\textbf{Technical $\leftrightarrow$ Human-centered}: Technical performance directly shapes user trust and experience, while human feedback and usage patterns influence technical effectiveness. In healthcare deployments, even systems with >95\% diagnostic accuracy failed when clinicians lacked calibrated trust, leading them to either over-rely on or dismiss system recommendations.

\textbf{Technical $\leftrightarrow$ Temporal}: Technical design choices determine long-term stability and adaptability, while temporal patterns reveal technical strengths and weaknesses. Financial AI systems that excelled in optimization tasks based on historical market data
but lacked robust adaptation mechanisms showed significant performance degradation during market volatility.

\textbf{Technical $\leftrightarrow$ Contextual}: Technical capabilities define what's possible within domain constraints, while contextual factors establish requirements and limitations for technical approaches. In retail, technically efficient systems that failed to align with customer emotional expectations damaged brand perception and reduced sales.

\textbf{Human-centered $\leftrightarrow$ Temporal}: User trust and mental models evolve over time, while system predictability and stability shape user expectations and behaviors. Trust calibration issues often compound over time, with initial minor discrepancies evolving into significant usage problems.

\textbf{Human-centered $\leftrightarrow$ Contextual}: Human experiences are shaped by organizational context and workflow fit, while user behavior determines how contextual value is realized. Systems that failed to integrate smoothly into established workflows were rejected despite strong technical performance.

\textbf{Temporal $\leftrightarrow$ Contextual}: Temporal adaptation must align with changing regulatory and business environments, while contextual factors determine what kinds of adaptation are valuable or permissible. Systems that couldn't adapt to seasonal retail patterns or evolving healthcare guidelines quickly lost their initial value.

These interdependencies may explain why single-dimension evaluation approaches often fail to predict real-world outcomes. Systems optimized solely for technical metrics may create unforeseen problems in other dimensions that only become apparent during deployment.
\subsection{Framework Implementation Across Domains}
To make our framework practical across different sectors, we propose an implementation approach that balances standardization with domain specificity. At the center lies a minimal core evaluation set, one foundational metric per axis, sufficient to provide an initial balanced assessment. From this foundation, each domain extends the framework with specialized metrics that address sector-specific needs while maintaining awareness of cross-dimensional effects.

Healthcare implementations tend to prioritize technical accuracy and human trust, financial services extend further on regulatory compliance and temporal stability, while retail emphasizes human satisfaction and contextual business metrics.

The implementation process should explicitly track interdimensional effects. For example, when improving diagnostic accuracy in healthcare, teams should simultaneously monitor changes in trust calibration, alert fatigue, and workflow integration to capture ripple effects across the system.

This structure supports both cross-domain comparability and contextual depth. It also enables scalable evaluation that is minimum viable testing at early stages, and deep analysis before full deployment.

\subsection{Metric Interaction Formalism}
To formalize our approach to balancing multiple metrics, we define a system's real-world effectiveness score. Let $T$, $H$, $R$, and $C$ be normalized scores (0–1) for technical, human-centered, temporal, and contextual metrics respectively. We define a system's real-world effectiveness score $U$ as:

\[
U = w_T T + w_H H + w_R R + w_C C \quad \text{where} \quad \sum w_i = 1
\]

In current practice, $w_T \approx 1$ and all others are near zero, implicitly treating technical success as a proxy for overall value. But evidence from~\cref{sec:meta-analysis} and~\cref{sec:case-studies} shows this fails to predict actual outcomes.

We argue that deployment-critical use cases (e.g., clinical decision support, financial compliance) require a \textbf{balanced weighting} scheme. For instance, $(w_T, w_H, w_R, w_C) = (0.3, 0.25, 0.2, 0.25)$, with calibration dependent on risk tolerance and domain complexity.

This formalism clarifies trade-offs and helps guide decisions during evaluation design. A system scoring 0.9 on $T$ but 0.2 on $H$ and $R$ may appear strong under conventional evaluation, yet would score only 0.55 under our balanced framework, correctly flagging potential deployment issues.

Beyond simple weighted combinations, advanced implementations should also consider interaction effects between dimensions. For example, the combination of low trust calibration $H$ with high adaptation rate $R$ can create particularly problematic outcomes in financial systems responding to market volatility.

Organizations can implement this framework through a phased approach that balances thoroughness with practical constraints.\cref{sec:pratice} provides a practical implementation example. 

\subsection{Anticipating and Addressing Counterarguments}

We anticipate four common objections to expanding evaluation beyond technical metrics:

\textbf{``Human-centered metrics are too subjective to trust.''} \textit{Response:} Subjectivity is not noise. It reflects user experience, which governs adoption and safety. Instruments like TrAAIT, NASA-TLX, and trust calibration curves have shown strong correlation with deployment outcomes. These can be standardized, validated, and benchmarked.

\textbf{``Safety and governance are outside the research scope; leave them to regulators.''} \textit{Response:} Safety, like accuracy, is an engineering problem first. Early design decisions shape emergent behaviors and risk profiles. Deferring evaluation until regulation is reactive and dangerous especially for rapidly evolving systems.

\textbf{``More metrics slow development and hinder innovation.''} \textit{Response:} Our framework is modular and scalable. Core metrics can be collected with minimal overhead, while deeper evaluation can be phased in during later stages. In fact, catching trust or adaptation issues early often \textit{accelerates} deployment by reducing backtracking.

\textbf{``One-size-fits-all frameworks can’t handle domain differences.''} \textit{Response:} That’s why our quadrant is extensible. The inner core supports cross-domain baselining, while outer layers incorporate sector-specific requirements. This mirrors how safety standards operate in aerospace vs. pharmaceuticals; different in detail, unified in structure.

This framework is not just a theoretical construct, it is designed for implementation. The next section provides concrete recommendations for researchers, industry practitioners, and policymakers to adopt, extend, and apply the model in practice.

\section{Recommendations and Research Agenda}
Addressing the measurement imbalance in agentic AI requires coordinated action across research, deployment, and governance communities. We propose the following priorities:

\subsection{Research Community}
\textbf{Develop and validate} human-centered evaluation instruments and temporal metrics across domains.

\textbf{Design cross-domain benchmarks} that integrate human dimensions, safety, and long-term performance beyond task correctness.

\textbf{Create lightweight evaluation toolkits} based on our quadrant framework, while incentivizing multidimensional reporting in publications through structured requirements.

\textbf{Bridge disciplines} by fostering collaboration between technical AI fields and HCI, safety engineering, and organizational science.

\subsection{Industry Practitioners}
\textbf{Implement comprehensive pre-deployment evaluations} across all four axes (technical, human, temporal, contextual) and track longitudinal agent behavior.

\textbf{Adopt trust-focused approaches} by measuring user interpretation and calibration early, while integrating staged evaluation into existing workflows.

\textbf{Enhance evaluation quality} by including domain experts in metric design and transparently reporting limitations.

\subsection{Policymakers}
\textbf{Mandate and standardize} human-centered metrics and longitudinal tracking for high-stakes AI applications.

\textbf{Fund open-source evaluation suites} aligned with the quadrant model, while coordinating cross-sector guidelines that balance harmonization with local context.

\textbf{Ensure accountability} through multidimensional reporting requirements for public-sector deployments, while creating regulatory ``safe harbors'' for testing new evaluation methods.

Evaluation forms the foundation for trust, effectiveness, and accountability in AI systems. This agenda offers a path toward more comprehensive and deployment-aligned assessment of agentic AI.

\section{Conclusion and Call to Action}
This paper has argued that current evaluation frameworks for agentic AI systems are dangerously imbalanced, privileging technical metrics while neglecting the human, temporal, and contextual dimensions that determine real-world value. Through a meta-analysis of 84 publications, cross-domain case studies, and a unifying framework, we have shown that benchmark success does not guarantee deployment success. As agents move from labs to high-stakes settings, our instruments of evaluation must evolve.

Looking ahead, the next generation of evaluation must address several urgent questions:
\begin{itemize}
    \item How can we standardize trust, collaboration, and workflow-fit metrics across domains without sacrificing contextual nuance?
    \item What are the minimal viable longitudinal tests that predict system degradation, adaptation, or failure modes before full-scale deployment?
    \item Can we design benchmarks that reflect not only agent intelligence but also their ability to operate safely, fairly, and sustainably in complex ecosystems?
\end{itemize}

We invite the community to help build the next era of agentic AI evaluation---one that reflects how systems succeed in the world, not just on the benchmark.

\begin{ack}

\end{ack}


\medskip

{
\small

\bibliography{neurips_2025.bib}  
\bibliographystyle{abbrv}


}


\appendix

\section{ Appendices }\label{app:A}
\subsection{Coding Schema for Meta-Analysis}\label{sec:codebook}
To characterize the evaluation focus of recent agentic AI literature across four key metric dimensions: technical performance, human-centered evaluation, safety/governance, and economic impact. Each publication was coded based on whether it included at least one metric or instrument aligned with each dimension. Both peer-reviewed academic publications and industry white papers were eligible. Papers were included in the review if they met all of the following:
\begin{itemize}
    \item Timeframe: Published between January 2023 and April 2025
    \item System Type: Described evaluation of an agentic AI system (LLM-based agents, multi-agent systems, autonomous decision-makers, or LLM tool users)
    \item Evaluation Evidence: Included at least one stated metric or evaluation result related to agent/system performance, user interaction, safety, or deployment outcomes
\end{itemize}

Each paper was coded using a binary vector [T, H, S, E], corresponding to~\cref{tab:code-dim}.
\begin{table*}[h!]
    \centering
    \small
    \begin{tabular}{>{\raggedright\arraybackslash}m{1cm} >{\raggedright\arraybackslash}m{3cm} > {\raggedright\arraybackslash}m{8.5cm}} 
\toprule
\textbf{Code} & \textbf{Dimension} & \textbf{Criteria for Inclusion} \\
\midrule
T  & Technical Performance & Includes any of the following: task success (e.g., Pass@k, accuracy), latency, resource usage (token count, memory), structural alignment (e.g., Tool F1, Node F1), robustness to noise or adversarial examples in narrow performance settings \\
H    & Human-Centered Evaluation & Includes user trust (surveys, retention, calibration), usability or workload measures (e.g., NASA-TLX), collaboration effectiveness, workflow integration, simulated human ratings (e.g., Agent-as-a-Judge), satisfaction/NPS, human-in-the-loop A/B tests \\
S   & Safety \& Governance & Includes adversarial robustness, value alignment, error recovery, failure rate analysis, regulatory compliance metrics, explainability metrics, fairness audits, or code-based autonomy scoring (e.g., orchestration code inspection)\\
E   & Economic \& Business Impact & Includes cost savings, ROI, time savings, process efficiency, conversion rate improvement, customer lifetime value, productivity uplift, or KPIs tied to organizational or business outcomes\\
Q   & Evaluation Quality  & Includes detailed methodology description, limitations acknowledgment, multiple evaluation methods, reproducibility information, or statistical validation of results\\
\bottomrule
\end{tabular}
\caption{Coding Dimensions and Definitions.}
\label{tab:code-dim}
\end{table*}
Papers were independently reviewed twice with a structured evaluation checklist (see below). Any ambiguity were resolved by a third reviewer. Full data and coded records are available in the accompanying CSV file.

\textbf{Metric Inclusion Checklist}

For each publication, the following yes/no questions were used:
\begin{itemize}
    \item T1: Does the paper report a performance metric on a specific task or benchmark?
    \item T2: Is any computational or latency efficiency reported?
    \item T3: Does the paper compare performance across different model versions, sizes, or against baseline systems?
    \item H1: Are human users (or proxies) involved in the evaluation?
    \item H2: Are trust, satisfaction, collaboration, or user behavior metrics reported?
    \item H3: Does the paper measure learnability, ease of use, or cognitive load aspects of the agent?
    \item S1: Are safety, failure rates, or edge case behaviors explicitly measured?
    \item S2: Are governance, compliance, or alignment frameworks included?
    \item S3: Are bias, fairness, ethical considerations, or representational harms measured or discussed with metrics?
    \item E1: Is business impact, ROI, or operational efficiency tracked?
    \item E2: Are KPIs from real-world deployment or simulations discussed?
    \item E3: Is scalability, cost-effectiveness at scale, or long-term economic viability evaluated?
    \item Q1: Is the evaluation methodology described in sufficient detail for reproducibility?
    \item Q2: Does the paper explicitly acknowledge limitations of the evaluation approach?
    \item Q3: Does the evaluation use multiple, complementary methods (triangulation) to validate findings?
\end{itemize}
If any question within a category was answered ``yes'' that dimension received a 1. If metrics simulate human experience (e.g., using an LLM to approximate user satisfaction), the paper was still marked H=1, with a flag for ``simulated human'' noted separately. Studies reporting only unstructured user impressions were excluded unless accompanied by coded instruments, clear KPIs, or quantifiable outcomes. Papers focused on emergent behavior or coordination were marked T=1, and S=1 if failure modes or robustness were analyzed.
\begin{figure}
    \centering
    \includegraphics[width=0.75\linewidth]{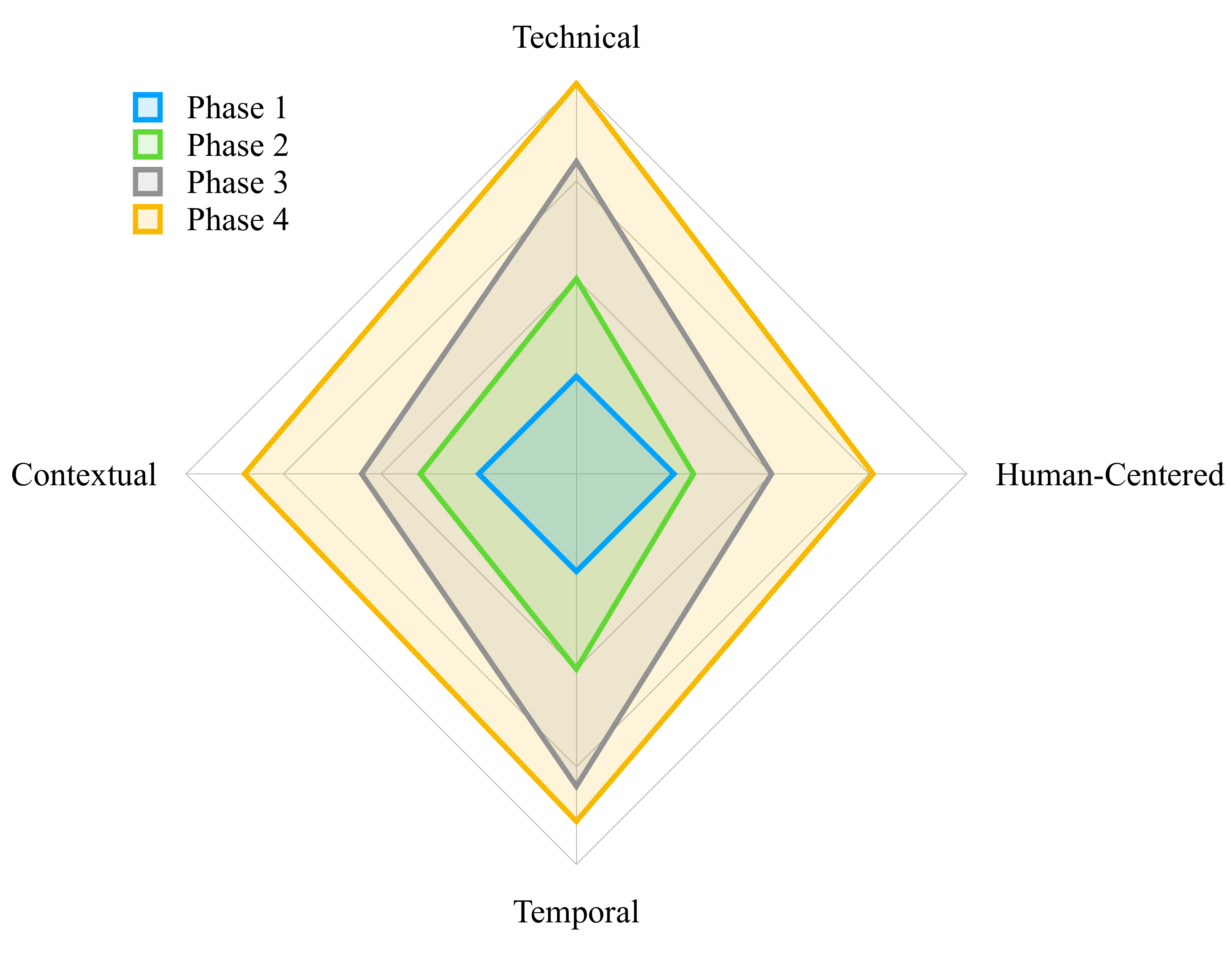}
    \caption{Framework implementation through a phased approach.}
    \label{fig:axis}
\end{figure}
\subsection{Practical Implementation Process}\label{sec:pratice}
Organizations can implement this framework through a phased approach that balances thoroughness with practical constraints:
\begin{itemize}
    \item Baseline Assessment~(phase 1): Establish core metrics across all four dimensions.
    \begin{itemize}
        \item Technical: Benchmark task completion and accuracy
        \item Human: Conduct initial trust calibration studies
        \item Temporal: Measure baseline stability in controlled conditions
        \item Contextual: Assess initial workflow fit and compliance requirements
    \end{itemize}
    \item Domain Adaptation~(phase 2): Select and calibrate domain-specific extensions
    \begin{itemize}
        \item Identify sector-specific validation instruments
        \item Define interdependence monitoring approach
        \item Establish evaluation thresholds based on deployment criticality
    \end{itemize}
    \item Pilot Evaluation~(phase 3): Apply the framework to controlled deployments
    \begin{itemize}
        \item Collect longitudinal data across all dimensions
        \item Monitor cross-dimensional effects
        \item Adjust weights based on observed interdependencies
    \end{itemize}
    \item Full Integration~(ongoing): Incorporate into development cycles and deployment decisions
    \begin{itemize}
        \item Maintain dimensional balance throughout scaling
        \item Regular reassessment as system and context evolve
        \item Proactive monitoring of emerging interdependence effects
    \end{itemize}
\end{itemize}
This phased approach balances comprehensiveness with real-world constraints, allowing organizations to begin shifting toward balanced evaluation without disrupting innovation cycles~(\cref{fig:axis}).

\subsection{Metric Type Distribution}\label{sec:dist}
\begin{figure}[h]
    \centering
    \includegraphics[width=0.9\linewidth]{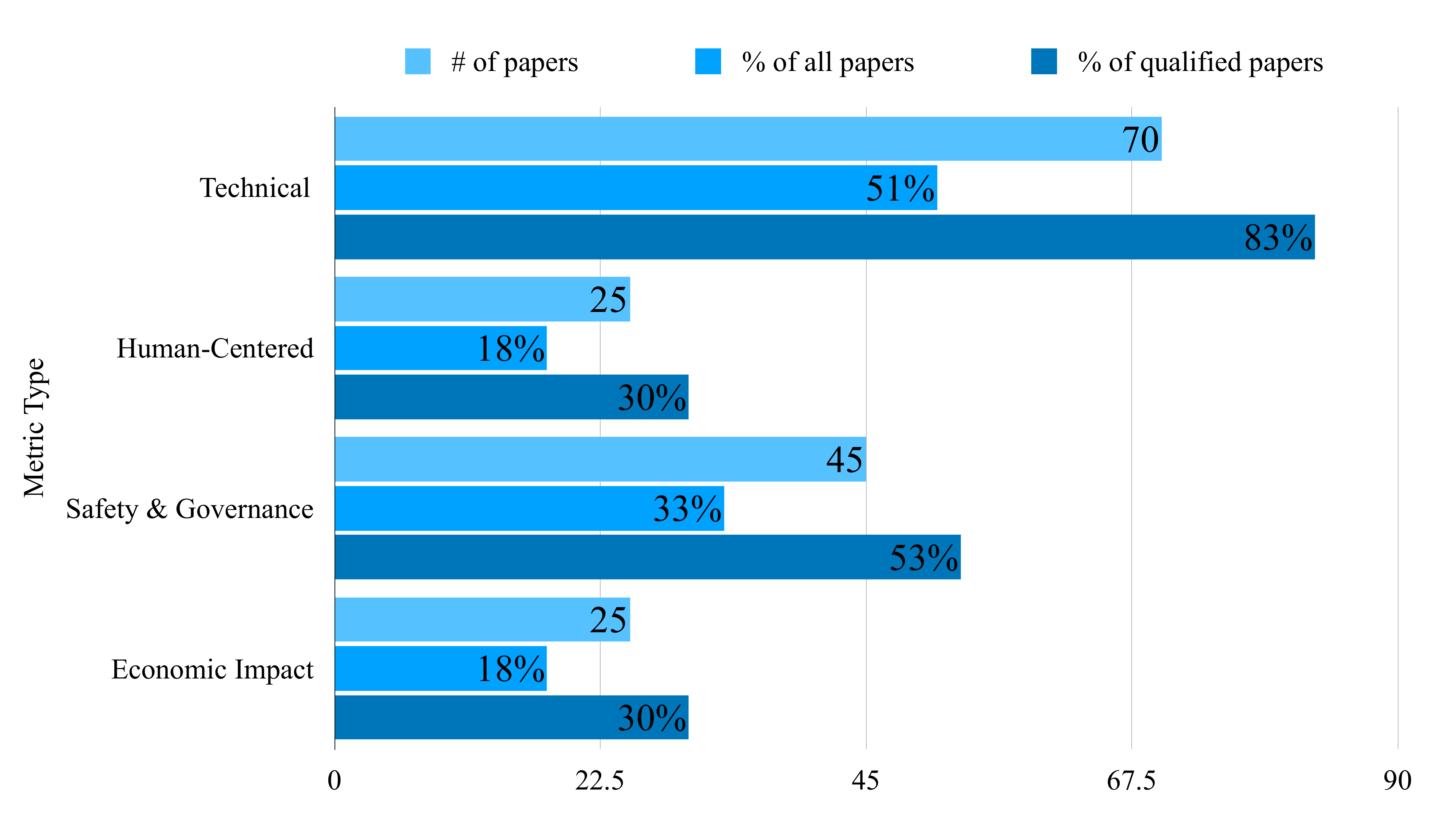}
    \caption{Distribution of evaluation metric types across agentic AI papers ($N=138$).}
    \label{fig:dist-chart}
\end{figure}

\cref{fig:dist-chart} displays three measures for each metric type: absolute number of papers, percentage of all examined papers, and percentage of qualified papers that met quality thresholds. Note that individual papers often employed multiple evaluation approaches.

\subsection{Terms and Definitions}\label{sec:terms}
\textbf{Agentic AI systems}: Systems characterized by: (1) goal-directed behavior with the ability to decompose complex objectives into manageable subtasks; (2) environmental awareness and adaptability to changing conditions; (3) tool utilization, where agents strategically leverage external resources to accomplish tasks; and (4) autonomous decision-making with limited human intervention.

\textbf{Measurement imbalance}: The systemic bias in evaluation frameworks that privilege easily quantifiable technical metrics while neglecting dimensions critical to real-world deployment success; especially human-centered factors, temporal stability, and contextual fit. This creates a fundamental misalignment between what we measure and what determines system value.

\textbf{Technical metrics}: Measures discrete system performance on well-defined tasks.
\begin{itemize}
    \item Task success (e.g., Pass@k, accuracy)
    \item Latency, resource usage (token count, memory)
    \item Structural alignment (e.g., Tool F1, Node F1)
    \item Robustness to noise or adversarial examples in narrow performance settings
    \item Computational or latency efficiency
    \item Performance comparisons across different model versions, sizes, or against baseline systems
\end{itemize}

\textbf{Human-centered metrics}: Captures how users experience, interpret, and adapt to the system. 
\begin{itemize}
    \item User trust (surveys, retention, calibration)
    \item Usability or workload measures (e.g., NASA-TLX)
    \item Collaboration effectiveness, workflow integration
    \item simulated human ratings (e.g., Agent-as-a-Judge)
    \item Satisfaction/NPS, human-in-the-loop A/B tests
    \item Learnability, ease of use, or cognitive load aspects
\end{itemize}

\textbf{Safety \& governance metrics}: Evaluates alignment with safety requirements and governance standards.
\begin{itemize}
    \item Adversarial robustness, value alignment, error recovery
    \item Failure rate analysis, regulatory compliance metrics
    \item Explainability metrics, fairness audits
    \item Code-based autonomy scoring (e.g., orchestration code inspection)
    \item Bias, fairness, ethical considerations, or representational harms measurements
\end{itemize}

\textbf{Economic \& business impact metrics}: Assesses financial and operational value creation.
\begin{itemize}
    \item Cost savings, ROI, time savings, process efficiency
    \item Conversion rate improvement, customer lifetime value
    \item Productivity uplift, or KPIs tied to organizational or business outcomes
    \item KPIs from real-world deployment or simulations
    \item Scalability, cost-effectiveness at scale, or long-term economic viability
\end{itemize}


\newpage

\end{document}